\documentclass{article}
\usepackage{amsfonts}
\usepackage{epsfig}
\usepackage{graphicx}
\usepackage{amssymb}

\title{Quantized Bertrand models on two-dimensional spaces of constant curvature}
\author{Agnieszka Martens\\
Helena Chodkowska University of Technology and Economics\\
Jutrzenki 135, 02-231 Warszawa, Poland\\
}

\begin{document}

\maketitle
\begin{abstract}
Discussed are Bertrand systems on two-dimensional  spaces of constant curvature, i.e.  sphere and pseudosphere (Lobachevsky space). These models often involve applying quantization techniques, such as those based on the Laplace-Beltrami operator, to the classical models and analyzing the resulting quantum behavior. 
\end{abstract}

\section{Introduction}

Considered are some quantization problems of what we used to call in \cite{m6} a test rigid body.
As usual, when quantizing systems in Riemannian configuration spaces, we use the Hilbert space $L^{2}(Q, \mu)$ of square-integrable functions,
where $\mu$ is the usual Riemannian measure,
\[
d\mu(q)=\sqrt{\left|\det[G_{ij}]\right|}dq^{1}\cdots
dq^{f},
\]
where $f$ denotes the number of degrees of freedom, i.e., $f=\dim
Q$ ($Q$ configuration space). For simplicity the square-root expression will be 
denoted by $\sqrt{\left|G\right|}$, where $G$ is the metric tensor field on $Q$. On the sphere and pseudosphere we have, respectively
\[
d\mu(r,\varphi, \psi)=R^{2} \sin^{2}\frac{r}{R}\sin\varphi dr d\varphi d  \psi ,
\]
\[
d\mu(r,\varphi, \psi)=R^{2} \sinh^{2} \frac{r}{R}\sin\varphi dr d\varphi d  \psi .
\]

The scalar product of wave amplitudes is given by 
\[
<\Psi_{1}|\Psi_{2}>=\int
\overline{\Psi}_{1}(q)\Psi_{2}(q)d\mu(q).
\]
The Hamilton (energy) operator is as follows
\[
\hat{H}=\hat{T}+V(r)=-\frac{\hbar ^{2}}{2m}\Delta +V(r),
\]
where the potential energy $V(r)$  is spherically symmetric and $ \Delta$ is the Laplace-Beltrami operator corresponding to
$G$,
\[
\Delta=\frac{1}{\sqrt{|G|}}\sum_{i,j}\partial_{i}
\sqrt{|G|}G^{ij}\partial_{j}=
G^{ij}\nabla_{i}\nabla_{j},
\]
where $\nabla$ denotes the Levi-Civita
covariant differentiation in the $G$-sense. 

The Laplace-Beltrami operator depending on the considered manifold is as follows \cite{m6}

\begin{itemize}
\item[$(i)$] sphere:
\begin{eqnarray}
\Delta & =&\frac{\partial ^{2} }{\partial r ^{2}}+\frac{1}{R}
\cot\frac{r}{R}\frac{\partial }{\partial r }-\frac{2\cos\frac{r}{R}}{R ^{2}\sin
^{2}\frac{r}{R}}\frac{\partial ^{2}}{\partial \varphi \partial \psi} \nonumber \\
&+&\frac{mR^{2}\sin
^{2}\frac{r}{R}+I\cos
^{2}\frac{r}{R}}{IR^{2}\sin
^{2}\frac{r}{R}}\frac{\partial ^{2} }{\partial \psi ^{2}}+\frac{1}{R^{2}\sin
^{2}\frac{r}{R}}\frac{\partial ^{2} }{\partial \varphi ^{2}},\label{EQ4}
\end{eqnarray}
\end{itemize}

\begin{itemize}
\item[$(ii)$] pseudosphere:
\begin{eqnarray}
\Delta &=&\frac{\partial ^{2} }{\partial r ^{2}}+\frac{1}{R}
\coth\frac{r}{R}\frac{\partial }{\partial r }-\frac{2\cosh\frac{r}{R}}{R ^{2}\sinh
^{2}\frac{r}{R}}\frac{\partial ^{2}}{\partial \varphi \partial \psi} \nonumber \\
&+& \frac{ mR^{2}\sinh
^{2}\frac{r}{R}+I\cosh
^{2}\frac{r}{R}}{IR^{2}\sinh
^{2}\frac{r}{R}}\frac{\partial ^{2} }{\partial \psi ^{2}}+\frac{1}{R^{2}\sinh
^{2}\frac{r}{R}}\frac{\partial ^{2} }{\partial \varphi ^{2}}.\label{EQ5a}
\end{eqnarray}
\end{itemize}

Separable solutions of the stationary Schr\"{o}dinger equation $\hat{H}\Psi =E\Psi$
have the form:
\begin{equation}\label{EQ6}
\Psi(r, \varphi, \psi)=f_{r}(r)f_{\varphi} (\varphi)f_{\psi} (\psi ).
\end{equation}
More convenient for our calculations is to use the variable $\vartheta=r/R$ 
\begin{equation}\label{EQ6a}
\Psi(\vartheta, \varphi, \psi)=f_{\vartheta}(\vartheta)e^{in\varphi}e^{il\psi},
\end{equation}
where $n, l $ are integers.

The stationary Schr\"{o}dinger equation with an arbitrary potential
$V(\vartheta)$ leads after the standard separation procedure
to the following one-dimensional radial eigenequations:
\begin{itemize}
\item[$(i)$] sphere:
\[
\frac{d^{2}f_{\vartheta}(\vartheta)}{d\vartheta^{2}}+\cot\vartheta \frac{df_{\vartheta}(\vartheta)}{d\vartheta}-
\]
\[
\left(\frac{\left(\frac{m}{I} R^{2}\sin^{2}\vartheta+\cos^{2}\vartheta\right)n^{2}+l^{2}-2nl\cos\vartheta}{\sin^{2}\vartheta}-
\frac{2mR^{2}}{\hbar^{2}}(E-V(\vartheta))\right)\] 
\[
f_{\vartheta}(\vartheta)=0,
\]

\item[$(ii)$] pseudosphere:
\[
\frac{d^{2}f_{\vartheta}(\vartheta)}{d\vartheta^{2}}+\coth\vartheta \frac{df_{\vartheta}(\vartheta)}{d\vartheta}-
\]
\[
\left(\frac{\left(\pm\frac{ m}{I} R^{2}\sinh^{2}\vartheta+\cosh^{2}\vartheta\right)n^{2}+l^{2}-2nl\cosh\vartheta}{\sinh^{2}\vartheta}-
\frac{2mR^{2}}{\hbar^{2}}(E-V(\vartheta))\right)
\]
\[
f_{\vartheta}(\vartheta)=0.
\]

\end{itemize}

Considered is a special case, when the translational part of the
potential energy $V (\vartheta)$ has the Bertrand structure, i.e. with the "frozen"
rotations all orbits would be closed.

\section{Bertrand potentials}

Bertrand systems on two‑dimensional spaces of constant curvature are the natural generalization of classical Bertrand systems from flat Euclidean space to curved geometries. There are two Bertrand potentials in Euclidean space for which all bounded orbits are closed: harmonic oscillator and attractive Coulomb problem. 

The one-dimensional radial eigenequations may be solved only when the explicit form of potential is specified:

\begin{itemize}
\item[$(a)$] oscillatory potentials:
\begin{equation}\label{b88}
V(r)=\frac{\gamma}{2}R^{2}{\rm \tan}^{2}\frac{r}{R},
\end{equation}

\item[$(b)$]  Kepler-Coulomb potentials:
\begin{equation}\label{c88}
V(r)=-\frac{\alpha}{R}{\rm \cot}\frac{r}{R}.
\end{equation}
\end{itemize}
Obviously, with the spherical topology also the geodetic problem belongs here:
\begin{itemize}
\item[$(c)$] $V(r)=0$, i.e., (in a sense) the special case of
$(a)$ or $(b)$ when $\gamma=0$, $\alpha=0$.
\end{itemize}
There is an obvious correspondence with the flat-space Bertrand
problem; it is suggested by the very asymptotics for $r\approx 0$,
i.e.,
\[
V(r)\approx\frac{\gamma}{2}r^{2}, \qquad V(r)\approx -\frac{\alpha}{r}.
\]
Obviously, this is a rough argument, but it may be shown
\cite{sx} that there exists a rigorous isomorphism
based on the projective geometry.

The resulting Schr\"{o}dinger equations should be rigorously solvable in terms of some standard
special functions. The most convenient way of solving them is to use the Sommerfeld
polynomial method \cite{m3}, \cite{m5}, \cite{m1}, \cite{m2}.
In this method the solutions are expressed by the usual or confluent Riemann $P$-functions. 
They are deeply related to the hypergeometric functions (respectively usual $F$ or confluent $F_{1}$). If the usual
convergence demands are imposed, then the hypergeometric functions become polynomials 
and our solutions are expressed by elementary functions. At the same time the energy 
levels and separation constants are expressed by the eigenvalues of the corresponding
operators. There exists some special class of potentials to which the Sommerfeld polynomial
method is applicable. The restriction to solutions expressible in terms of Riemann 
$P$-functions is reasonable, because this class of functions is well investigated and 
many special functions used in physics may be expressed by them.

\subsection{Oscillatory potentials}

\begin{itemize}
\item[$(i)$] sphere:
\end{itemize}

\[
V(r)=\frac{\gamma}{2}R^{2}\tan^{2}\frac{r}{R}.
\]
Here obtain the energy levels $E$  and the function $f_{r}(r)$ in the form:\\
\begin{eqnarray}\label{EQ8}
E&=&\frac{1}{2}\hbar\Omega\left(\left(2k+1+|n-l|+\sqrt{(n+l)^{2}+
\frac{\gamma mR^{4}}{\hbar^{2}}}\right)^{2}\right. \nonumber \\ 
&+ & \left. 4n^{2}\left(\frac{m}{I}R^{2}-1\right)-\frac{4\gamma mR^{4}}{\hbar^{2}}-1\right),
\end{eqnarray}
where $\Omega=\hbar \omega / 4mR^{2}$, $\omega=\sqrt{\gamma/m}$ and $k=0,1,... \ $. 
\begin{equation}
f_{r}(r)=\left(\cos \frac{r}{R}\right)^{\kappa}\left(\sin \frac{r}{R}\right)^{\nu}  F\left(-k,k+1+\kappa+\nu;1+\kappa;\cos^{2} \frac{r}{R}\right),
\end{equation}
where 
\[
\kappa=\sqrt{(n+l)^{2}+\frac{\gamma mR^{4}}{\hbar^{2}}}, \quad \nu=|n-l|.
\]
\begin{itemize}
\item[$(ii)$] pseudosphere:
\end{itemize}
We take the "harmonic oscillator" - type potential: 
\[
V(r)=\frac{\gamma}{2}R^{2}\tanh^{2}\frac{r}{R}, \quad \gamma > 0.
\]
\bigskip
We find the energy levels:
\begin{eqnarray}\label{EQ9a}
E&=&\frac{1}{2}\hbar\Omega\left(\left(2k+1+|n-l|+\sqrt{(n+l)^{2}+
\frac{\gamma mR^{4}}{\hbar^{2}}}\right)^{2}\right. \nonumber \\ 
&- & \left. 4n^{2}\left(\pm\frac{m}{I}R^{2}-1\right)-\frac{4\gamma mR^{4}}{\hbar^{2}}-1\right).
\end{eqnarray}
The function $f_{r}(r)$ is as follows:
\begin{equation}
f_{r}(r)=\left(\cosh \frac{r}{R}\right)^{\kappa}\left(\sinh \frac{r}{R}\right)^{\nu}  F\left(-k,k+1+\kappa+\nu;1+\kappa;\cosh^{2} \frac{r}{R}\right).
\end{equation}
We can notice that 
\[
\lim_{r \rightarrow \infty}\frac{\gamma}{2}R^{2}\tanh^{2}\frac{r}{R}=\frac{\gamma}{2}R^{2}.
\]
This is the upper bound of the potential $V$. Therefore, the formula (\ref{EQ9a}) is correct only for such quantum numbers that
\[
E < {\rm Sup} \ V = \frac{\gamma}{2} R^{2}.
\]
Above this threshold we are dealing with the continuous spectrum and the classically non-restricted motion.

\subsection{Kepler-Coulomb potentials}

\begin{itemize}
\item[$(i)$] sphere:
\end{itemize}
\begin{equation}\label{c88}
V(r)=-\frac{\alpha}{R}{\rm \cot}\frac{r}{R}.
\end{equation}
Here we obtain the energy
levels $E$ as follows:\\
\begin{equation}\label{EQ8}
E=-\frac{\hbar^{2}}{2mR^{2}}\left( (k+|n+l||n-l|) -\left( \left(\frac{m}{I}R^{2}+1 \right)n^{2}-4nl-2\alpha \right) \right),
\end{equation}
where $k=0,1,... \ $. And the function $f_{r}(r)$ in the form:
\begin{equation}
f_{r}(r)=\left(\sin \frac{r}{R}\right)^{\kappa}\left(\cos \frac{r}{R}\right)^{\nu}  F\left(-k,k+\kappa+\nu;1+\kappa;\sin \frac{r}{R}\right),
\end{equation}
where 
\[
\kappa=|n+l|, \quad \nu=|n-l|.
\]
\begin{itemize}
\item[$(ii)$] pseudosphere:
\end{itemize}
We take the "attractive Kepler-Coulomb" - type potentials: 
\begin{equation}\label{b133}
 V(r)=-\frac{\alpha}{R} {\rm \coth}\frac{r}{R}, \qquad \alpha>0.
\end{equation}
\bigskip
In this case the energy levels $E$ as as follows:
\begin{equation}\label{EQ9}
E=-\frac{\hbar^{2}}{2mR^{2}}\left( (k+|n+l||n-l|) +\left( \left(\pm \frac{m}{I}R^{2}+1 \right)n^{2}-4nl-2\alpha \right) \right).
\end{equation}
The function $f_{r}(r)$ has the form:
\begin{equation}
f_{r}(r)=\left(\sinh \frac{r}{R}\right)^{\kappa}\left(\cosh \frac{r}{R}\right)^{\nu}  F\left(-k,k+\kappa+\nu;1+\kappa;\sinh \frac{r}{R}\right).
\end{equation}

\section{Conclusions}

The considered quantized Bertrand models are completely non-degenerate because of 
the existence of three quantum numbers labelling the energy levels. They cannot be               
combined into a single quantum number. We can notice that: there is no total
quantum degeneracy, i.e., hyperintegrability, with respect to them and
the interaction between translational and rotational degrees of freedom completely removes degeneracy.


\begin{thebibliography}{9}


\bibitem{2} V. I. Arnold,  {\em Mathematical Methods of Classical Mechanics\/}, Springer Graduate Texts in Mechanics {\bf 60}, Springer Verlag, New York 1978.

\bibitem{m3}
A. Martens: {\em Rep.\ Math.\ Phys.\/}, {\bf 51}, 287-295 (2003).

\bibitem{m4}
A. Martens: {\em J.\ of Nonlinear Math.\ Phys.\/}, {\bf 11}, Supplement, 145-150 (2004).

\bibitem{m5}
A. Martens: {\em J.\ of Nonlinear Math.\ Phys.\/}, {\bf 11}, Supplement, 151-156 (2004).


\bibitem{m1}
A. Martens: {\em Rep.\ Math.\ Phys.\/}, {\bf 62}, 145-155 (2008).

\bibitem{m2}
A. Martens, J. J. S\l awianowski: {\em Acta\ Phys.\ Pol. B\/}, {\bf 41} 1847-1880 (2010).


\bibitem{m6}
A. Martens,  {\em Rep. Math. Phys.} {\bf 71}, 381  (2013).

\bibitem{m7}
A. Martens,  {\em Acta\ Phys.\ Pol. B\/}  {\bf 46},  843  (2015).

\bibitem{JJS} J.J. S\l awianowski, {\em Geometry of Phase Spaces},  John Wiley \& Sons, Chichester, New York, Brisbane, Toronto, Singapore, PWN-Polish Scientific Publishers, Warszawa 1991. 


\bibitem{sx} 
J. J. S\l awianowski: {\em Rep.\ Math.\ Phys.\/}, {\bf 46}, 429-460 (2000). 

\end{thebibliography}
\end{document}